\documentclass{aa}

\usepackage[utf8]{inputenc}
\usepackage[T1]{fontenc}

\usepackage[english]{babel}
\usepackage{txfonts}
\usepackage{charter}
\usepackage{balance}
\usepackage{hyperref}
    \hypersetup{colorlinks=true,linkcolor=blue,citecolor=blue,urlcolor=blue}
    \makeatletter
    \renewcommand*\aa@pageof{, page \thepage{} of \pageref*{LastPage}}
    \makeatother
\usepackage{lipsum}

\usepackage{siunitx}

\usepackage{booktabs}
\usepackage[flushleft,para]{threeparttable}
\usepackage{multirow}
\usepackage{rotating}

\usepackage[nowarn]{glossaries}
    \makeglossaries

\newacronym{BIS}{BIS}{bisector inverse slope}

\newacronym{CB}{CB}{convective blueshift}
\newacronym{CCF}{CCF}{cross-correlation function}

\newacronym{DACE}{DACE}{Data \& Analysis Center for Exoplanets}
\newacronym{DRS}{DRS}{Data Reduction Software}

\newacronym{FAP}{FAP}{false alarm probability}
\newacronym{FWHM}{FWHM}{full-width at half-maximum}

\newacronym{GLS}{GLS}{Generalized Lomb-Scargle}

\newacronym{HARPS}{HARPS}{High Accuracy Radial velocity Planet Searcher}
\newacronym{HARPS-N}{HARPS-N}{HARPS-North}

\newacronym{LBL}{LBL}{line-by-line}

\newacronym{MLR}{MLR}{multi-linear regression}
\newacronym{MS}{MS}{main-sequence}

\newacronym{PBP}{PBP}{part-by-part}
\newacronym{PCA}{PCA}{principal component analysis}

\newacronym{RMS}{RMS}{root mean square}
\newacronym{RV}{RV}{radial velocity}

\newacronym{S/N}{S/N}{signal-to-noise ratio}
\newacronym{SME}{SME}{Spectroscopy Made Easy}

\newacronym{VALD}{VALD}{Vienna Atomic Line Database}

\usepackage{orcidlink}

\begin{document}

   \title{Measuring precise radial velocities on individual spectral lines}
   \subtitle{IV. Stellar activity correlation with line formation temperature\thanks{Based on observations made with the HARPS instrument on the ESO 3.6m Telescope at the La Silla Observatory, Chile.}}

   \author{K. Al Moulla \inst{1}\orcidlink{0000-0002-3212-5778}
      \and X. Dumusque  \inst{1}\orcidlink{0000-0002-9332-2011}
      \and M. Cretignier\inst{2}\orcidlink{0000-0002-2207-0750}}

   \institute{Observatoire Astronomique de l'Université de Genève, Chemin Pegasi 51, 1290 Versoix, Switzerland\\
   \email{khaled.almoulla@unige.ch}
         \and{Sub-department of Astrophysics, Department of Physics, University of Oxford, Oxford OX1 3RH, UK}
             }

   \date{Received 4 October 2023 / Accepted 12 December 2023}

\abstract
{Radial velocities (RVs) of stars contain both the Doppler reflex motion of potential planetary companions and the drowning and sometimes imitating effect of stellar activity. To separate the two, previous efforts have sought for proxys which only trace the activity signals, yet the sub-meter-per-second floor required for the detection of Earth-like planets remains difficult to break.}
{In this work, we analyze a sample of 12 G- to early M-type stars in order to investigate the feasibility of detecting a differential effect of stellar activity with photospheric depth, as traced by the spectral line-forming temperature, for observations with different sampling and noise levels.}
{We computed the average line formation temperature for each point in the observed wavelength grids using the spectral synthesis code \texttt{PySME}. The final line selection was curated to exclude blended and poorly synthesized lines. We thereafter computed the \acrfull*{CB} of the line cores of our master spectra (composed of the stacked individual spectra for each star). Finally, we extract RV time series for certain intervals of formation temperature using a template-matching approach.}
{We find the CB to follow a linear relation with the formation temperature of the line cores, and the CB slope to be steeper with increasing effective temperature. For the RV time series derived for different intervals of formation temperature, we find the RVs of line parts formed at higher temperatures, close to the spectral continuum, to be generally correlated with the $S$ index, and RVs of line parts formed at cooler temperatures, close to the spectral line cores, to be generally anti-correlated, especially for stars with low noise levels and significant variations over their magnetic cycles.}
{RVs of line parts formed in the coolest 25\% of the line-forming temperature range appear to be a strong tracer of stellar activity over the magnetic cycle for several stars. By detrending the total RV time series with a multi-linear combination of residuals of RVs measured at different temperature ranges and the $S$ index, the RV scatter can be decreased to a greater extent than with the $S$ index alone.}


\authorrunning{K. Al Moulla et al.}
\titlerunning{Measuring precise radial velocities on individual spectral lines. IV.}

\maketitle


\section{Introduction}\label{Sect:1}

Currently, stellar activity is one of the primary obstacles for detecting the \acrfull*{RV} signal of Earth-like exoplanets around solar-type stars. Solar variability has been thoroughly monitored \citep[e.g.][]{Phillips+2016,CollierCameron+2019,Lin+2022}, characterized \citep[e.g.,][]{Kjeldsen&Bedding1995,Lefebvre+2008,AlMoulla+2023} and simulated \citep[e.g.,][]{Boisse+2012,Dumusque+2014,Meunier+2015,Ervin+2022,Zhao&Dumusque2023} in Sun-as-a-star RV observations. Extracting our knowledge of the Sun to other stars, we have come to understand that activity phenomena, such as acoustic oscillations \citep[e.g.,][]{Kjeldsen+2005,Arentoft+2008,Chaplin+2019}, granulation \citep{Dumusque+2011a,Meunier&Lagrange2019} and active regions \citep[e.g.][]{Meunier+2010,Dumusque+2011b,Lovis+2011}, i.e., spots, faculae and plages, all produce surface velocity variations on the order of tens of meters-per-second, which drown out the eventual ${\sim}\SI{0.1}{\meter\per\second}$ semi-amplitude signal of an Earth analogue.

Previous efforts using stellar activity indicators, such as the \acrfull*{FWHM} and the \acrfull*{BIS} of the \acrlong*{CCF} \citep[CCF;][]{Queloz+2001,Pepe+2002} or the Mount Wilson $S$ index extracted from the \ion{Ca}{II} H and K lines \citep{Baliunas+1983}, have not been able to sufficiently mitigate the activity signal since these indicators are seldom fully correlated with the RVs \citep{Haywood+2016}. Recent studies \citep[e.g.,][]{Cretignier+2021,Cretignier+2022} have instead attempted to implement mitigating techniques at the spectral level, before the CCF is applied to extract the RVs, since it is now well established that stellar activity (primarily the inhibition of convective blueshift by magnetically active regions), affects spectral lines to different extents \citep[e.g.,][]{Reiners+2016,Dumusque2018}.

In this study, we investigate how stellar activity affects individual spectral lines parts formed at various temperature regimes in the stellar photosphere. We apply the methodology of \cite{AlMoulla+2022}, which studied this dependence for the Sun and $\alpha\,\mathrm{Cen}\,\mathrm{B}$ (HD 128621, which is also included in our sample to be analyzed on the long-term), and found that RVs of line parts formed in hotter parts of the photosphere, close to the spectral continuum, are more strongly correlated with the RV contribution of the inhibition of convective blueshift compared to line parts formed at cooler temperatures. Here, we extend the work to a sample of 12 late-type (G to early-M) dwarfs, in order to investigate whether similar effects can be seen in other spectral types.

In Sect.~\ref{Sect:2}, we describe the stellar sample and the available spectral data. In Sect.~\ref{Sect:3}, we describe the method used to compute the RV of selected line parts. In Sect.~\ref{Sect:4}, we analyze the correlations between stellar activity and formation temperature-binned RV signals. In Sect.~\ref{Sect:5}, we discuss our results and conclude.

\section{Data}\label{Sect:2}

\subsection{Spectral pre-treatment}\label{Sect:2.1}

The stellar sample consists of 12 G- to early M-type stars observed  with the HARPS\footnote{HARPS stands for \acrlong*{HARPS}.} spectrograph, mounted on the ESO 3.6m Telescope at the La Silla Observatory, Chile, which covers a spectral range from $\SI{378}{}$ to $\SI{691}{nm}$ and provides an instrumental resolution of $R\,{=}\,\num{115000}$. The targets have been frequently observed at relatively high \acrfull*{S/N} and exhibit clear $S$ index variations indicative of magnetic activity cycles over the observed time span (except for HD 211038 which appears to be quiet in recent years). They are listed in Table~\ref{Tab:01}, with their observational data, detailing the number of spectra per target, the median S/N, the span in $S$ index after smoothing with a 200-day rolling average (to be less affected by outliers and rotational modulation of the $S$ index), and the span in $\log{R'_{\mathrm{HK}}}$ converted from the smoothed $S$ index \citep{Noyes+1984}. $\log{R'_{\mathrm{HK}}}$ is only a re-scaling of the $S$ index, however, it also corrects for the photospheric contribution to the emission in the \ion{Ca}{II} H and K lines using the B{$-$}V color for each star (see Table~\ref{Tab:02}), permitting a better comparison between spectral types. We list the span in $\log{R'_{\mathrm{HK}}}$ in Table~\ref{Tab:01} to compare the relative variations within our stellar sample, however, we use the $S$ index for the remainder of our analysis since we will only consider correlations with the activity proxies and not their absolute values. Note that the listed numbers of spectra represent the number of retained and nightly-stacked spectra after processing with \texttt{YARARA} \citep{Cretignier+2021,Cretignier+2023}, during which a quality-based rejection could be made, and hence the numbers could differ compared to archival data. \texttt{YARARA} performs corrections of stellar activity, telluric absorption and instrumental systematics using mainly \acrfull*{PCA} applied on the residual of the spectral data matrix (residual flux versus time).

\texttt{YARARA} also provides a master spectrum computed as the inverse-variance weighted average of all individual spectra at each sampled wavelength point. The absolute RV shift of the master spectrum might lead to an arbitrary offset when template-matching with the individual spectra (for example, if the RV of the master spectrum is high, low-activity spectra will appear slightly more blueshifted relative to the master compared to a master constructed using only the quietest observations). However, since we are only interested in RV variations, this offset should not be of much importance. More importantly, the \texttt{YARARA} corrections work better at higher S/N which is why we opted to create master spectra using all available observations. These masters serve as both the references to compare the syntheses (Sect.~\ref{Sect:3.1}) with true observations, and to perform template-matching when extracting the temperature-binned RV time series (Sect.~\ref{Sect:3.4}). After the corrections, we only re-inject the corrected stellar activity of the individual spectra in order to isolate its effect at the spectral level.

\subsection{Stellar parameters}\label{Sect:2.2}

\begin{table}[t!]
\centering
\caption{Observational specifications relevant for our analysis. The columns specify the target name, spectral type, number of spectra, median S/N after nightly stacking, span (peak to peak difference) in $S$ index after smoothing with a 200-day rolling average (see Figs.~\ref{Fig:A1}--\ref{Fig:A12}), and span in $\log{R'_{\mathrm{HK}}}$ converted from the smoothed $S$ index.}
\begin{tabular*}{\linewidth}{l @{\extracolsep{\fill}} l @{\extracolsep{\fill}} S[table-format=3.0] @{\extracolsep{\fill}} S[table-format=3.0] @{\extracolsep{\fill}} S[table-format=1.3] @{\extracolsep{\fill}} S[table-format=1.3]}
\toprule
{Target}  & {Sp. type} & {$N_{\mathrm{spec}}$} & {S/N} & {${\Delta}S$} & {${\Delta}\log{R'_{\mathrm{HK}}}$} \\
\midrule
Gl 229    & M1V        &                    92 &   136 & 0.412         & 0.115                              \\
HD 36003  & K5V        &                    78 &   169 & 0.124         & 0.146                              \\
HD 85512  & K6V        &                   473 &   182 & 0.193         & 0.201                              \\
HD 154577 & K2.5V      &                   295 &   169 & 0.044         & 0.108                              \\
HD 215152 & K3V        &                   269 &   124 & 0.080         & 0.150                              \\
HD 211038 & G9.5IV     &                    68 &   232 & 0.003         & 0.013                              \\
HD 192310 & K2V        &                   323 &   309 & 0.052         & 0.149                              \\
HD 128621 & K1V        &                   292 &   583 & 0.054         & 0.163                              \\
HD 21693  & G9V        &                   194 &   128 & 0.035         & 0.141                              \\
HD 56274  & G7V        &                   146 &   145 & 0.013         & 0.068                              \\
HD 189567 & G2V        &                   221 &   239 & 0.010         & 0.050                              \\
HD 45184  & G2V        &                   166 &   243 & 0.017         & 0.091                              \\
\bottomrule
\end{tabular*}
\label{Tab:01}
\end{table}

\begin{sidewaystable*}
\centering
\caption{Parameters of the stellar sample. The columns indicate the target name, spectral type, B{$-$}V color, effective temperature, surface gravity, metallicity, projected rotational velocity, mass, radius, luminosity, and references. The stars are ordered by effective temperature. For HD 21693, the mass and radius are computed using the effective temperature, surface gravity and luminosity as described in the text.}
\begin{tabular*}{\textwidth}{l @{\extracolsep{\fill}} l @{\extracolsep{\fill}} S[table-format=1.2] @{\extracolsep{\fill}} S[table-format=4.0]@{\extracolsep{0pt}}@{$\,$}c@{$\,$}@{\extracolsep{0pt}}S[table-format=3.0] @{\extracolsep{\fill}} S[table-format=1.2]@{\extracolsep{0pt}}@{$\,$}c@{$\,$}@{\extracolsep{0pt}}S[table-format=1.2] @{\extracolsep{\fill}} S[table-format=-1.2]@{\extracolsep{0pt}}@{$\,$}c@{$\,$}@{\extracolsep{0pt}}S[table-format=1.2] @{\extracolsep{\fill}} S[table-format=1.3]@{\extracolsep{0pt}}@{$\,$}c@{$\,$}@{\extracolsep{0pt}}S[table-format=1.3] @{\extracolsep{\fill}} S[table-format=1.3]@{\extracolsep{0pt}}@{$\,$}c@{$\,$}@{\extracolsep{0pt}}l @{\extracolsep{\fill}} S[table-format=1.3]@{\extracolsep{0pt}}@{$\,$}c@{$\,$}@{\extracolsep{0pt}}l @{\extracolsep{\fill}} S[table-format=1.3]@{\extracolsep{0pt}}@{$\,$}c@{$\,$}@{\extracolsep{0pt}}l @{\extracolsep{\fill}} l}
\toprule
{Target} & {Sp. type} & {B$-$V} & \multicolumn{3}{c}{{$T_{\mathrm{eff}}$ [K]}} & \multicolumn{3}{c}{{$\log{g}$}} & \multicolumn{3}{c}{{[Fe/H]}} & \multicolumn{3}{c}{{$v\sin{i}$ [km/s]}} & \multicolumn{3}{c}{{$M_{\star}$ [$M_{\sun}$]}} & \multicolumn{3}{c}{{$R_{\star}$ [$R_{\sun}$]}} & \multicolumn{3}{c}{{$L_{\star}$ [$L_{\sun}$]}} & {References} \\
\midrule
   Gl 229 &        M1V &    1.48 &                     3733 & $\pm$ &  50 &        5.02 & $\pm$ & 0.25 &    -0.05 & $\pm$ & 0.25 &                 2.6 & $\pm$ &   1.0 &                       0.58 &      $\pm$ &                {-} &                       0.46 & $^{+}_{-}$ & $^{0.01}_{0.02}$ &                            &   {-} &      &      1,2,3,4 \\
 HD 36003 &        K5V &    1.14 &                     4647 & $\pm$ &  88 &        4.31 & $\pm$ & 0.21 &     -0.2 & $\pm$ & 0.06 &               2.063 & $\pm$ & 0.116 &                       0.75 &      $\pm$ &              0.008 &                        0.7 &      $\pm$ &             0.02 &                            &   {-} &      &        4,5,6 \\
 HD 85512 &        K6V &    1.18 &                     4715 & $\pm$ & 102 &        4.39 & $\pm$ & 0.28 &    -0.32 & $\pm$ & 0.03 &               2.194 & $\pm$ & 0.118 &                      0.712 & $^{+}_{-}$ & $^{0.008}_{0.031}$ &                       0.66 & $^{+}_{-}$ & $^{0.03}_{0.01}$ &                            &   {-} &      &        4,5,6 \\
HD 154577 &      K2.5V &    0.89 &                     4881 & $\pm$ &  44 &        4.61 & $\pm$ & 0.06 &    -0.56 & $\pm$ & 0.03 &                 0.2 & $\pm$ &   0.3 &                      0.688 & $^{+}_{-}$ & $^{0.028}_{0.026}$ &                       0.68 &      $\pm$ &             0.02 &                            &   {-} &      &          4,7 \\
HD 215152 &        K3V &    0.99 &                     4935 & $\pm$ &  76 &         4.4 & $\pm$ & 0.14 &     -0.1 & $\pm$ & 0.04 &               1.922 & $\pm$ & 0.059 &                       0.77 &      $\pm$ &              0.015 &                       0.73 &      $\pm$ &             0.02 &                            &   {-} &      &        5,6,8 \\
HD 211038 &     G9.5IV &    0.90 &                     5009 & $\pm$ &  44 &        3.87 & $\pm$ & 0.06 &    -0.13 & $\pm$ & 0.03 &                 2.7 & $\pm$ &   0.5 &                       1.15 & $^{+}_{-}$ & $^{0.034}_{0.038}$ &                       2.44 & $^{+}_{-}$ &  $^{0.1}_{0.07}$ &                            &   {-} &      &          4,7 \\
HD 192310 &        K2V &    0.91 &                     5080 & $\pm$ &  25 &        4.55 & $\pm$ & 0.03 &      0.0 & $\pm$ & 0.02 &                 2.0 & $\pm$ &   0.3 &                      0.816 & $^{+}_{-}$ & $^{0.024}_{0.013}$ &                       0.81 & $^{+}_{-}$ & $^{0.02}_{0.01}$ &                            &   {-} &      &          4,7 \\
HD 128621 &        K1V &    0.88 &                     5178 & $\pm$ &  22 &        4.56 & $\pm$ & 0.03 &     0.15 & $\pm$ & 0.01 &                 0.9 & $\pm$ &   0.3 &                      0.877 & $^{+}_{-}$ & $^{0.028}_{0.013}$ &                       0.88 &      $\pm$ &             0.02 &                            &   {-} &      &          4,7 \\
 HD 21693 &        G9V &    0.78 &                     5430 & $\pm$ &  26 &        4.37 & $\pm$ & 0.04 &      0.0 & $\pm$ & 0.02 &               1.988 & $\pm$ & 0.289 &                      0.676 &      $\pm$ &              0.028 &                      0.888 &      $\pm$ &            0.017 &                      0.618 & $\pm$ & 0.02 &          5,6 \\
 HD 56274 &        G7V &    0.57 &                     5681 & $\pm$ &  44 &        4.47 & $\pm$ & 0.06 &     -0.5 & $\pm$ & 0.03 &                 1.7 & $\pm$ &   0.5 &                            &        {-} &                    &                            &        {-} &                  &                            &   {-} &      &            7 \\
HD 189567 &        G2V &    0.64 &                     5713 & $\pm$ &  25 &        4.45 & $\pm$ & 0.03 &    -0.18 & $\pm$ & 0.02 &                 1.2 & $\pm$ &   0.3 &                      0.922 & $^{+}_{-}$ &  $^{0.029}_{0.03}$ &                       1.04 &      $\pm$ &             0.03 &                            &   {-} &      &          4,7 \\
 HD 45184 &        G2V &    0.62 &                     5810 & $\pm$ &  44 &        4.37 & $\pm$ & 0.06 &     0.03 & $\pm$ & 0.03 &                 2.5 & $\pm$ &   0.5 &                      1.007 & $^{+}_{-}$ & $^{0.024}_{0.022}$ &                       1.08 & $^{+}_{-}$ & $^{0.04}_{0.03}$ &                            &   {-} &      &          4,7 \\
\bottomrule
\end{tabular*}
\begin{flushleft}
\small
\textbf{References.} 1: \citealt{Kuznetsov+2019}, 2: \citealt{Hojjatpanah+2020}, 3: \citealt{Tuomi+2014}, 4: \citealt{Takeda+2007}, 5: \citealt{Sousa+2008}, 6: \citealt{Soto&Jenkins2018}, 7: \citealt{Valenti&Fischer2005}, 8: \citealt{Delisle+2018}
\end{flushleft}
\label{Tab:02}
\end{sidewaystable*}

The adopted stellar parameters are listed in Table~\ref{Tab:02}. The spectral types and B{$-$}V colors are retrieved from SIMBAD\footnote{\url{https://simbad.u-strasbg.fr}} \citep{Wenger+2000}. The effective temperatures, $T_{\mathrm{eff}}$, surface gravities, $\log{g}$, metallicities, [Fe/H], projected rotational velocities, $v\sin{i}$, masses, $M$, and radii, $R$, are retrieved from the references in the table foot. For HD 21693, mass and radius could not be found in the literature, and were instead computed from the luminosity, $L$, using the following formulas from \citet{Valenti&Fischer2005}:
\begin{subequations}
\begin{align}
&\frac{R}{R_{\sun}} = \sqrt{\frac{L}{L_{\sun}}}\left(\frac{T_{\mathrm{eff},\sun}}{T_{\mathrm{eff}}}\right)^{2} \label{Eq:1a} \\
&\sigma_{R} = R\sqrt{\left(\frac{\sigma_{L}}{2L}\right)^{2} + \left(\frac{2\sigma_{T_{\mathrm{eff}}}}{T_{\mathrm{eff}}}\right)^{2}} \label{Eq:1b}
\end{align}
\end{subequations}
\begin{subequations}
\begin{align}
&\frac{M}{M_{\sun}} = \left(\frac{R}{R_{\sun}}\right)^{2}10^{\log{g} - \log{g}_{\sun}} \label{Eq:2a} \\
&\sigma_{M} = M\sqrt{\left(\frac{\sigma_{\log{g}}}{\log{e}}\right)^{2} + \left(\frac{2\sigma_{R}}{R}\right)^{2}} \label{Eq:2b}
\end{align}
\end{subequations}
where quantities with subscript $\sun$ denote solar values, and $\sigma_{M}$, $\sigma_{R}$ and $\sigma_{L}$ are the uncertainties of the mass, radius and luminosity, respectively. It should be noted that the masses and radii are solely used for the computation of the gravitational redshift (see Sect.~\ref{Sect:3.3}) and are not needed for the subsequent computations of the temperature-dependent RVs.

\section{Methods}\label{Sect:3}

One of the primary ways of investigating the effects of stellar activity at the spectral level is by extracting its differential effect on spectral lines as seen in \acrfull*{LBL} RVs \citep{Dumusque2018,Cretignier+2020,Artigau+2022,Lafarga+2023}. In order to further trace the origin of activity, one can instead of considering entire spectral lines (or spectral regions bounded by two local maxima as in the case of M dwarfs where spectral lines are everywhere), consider various line or CCF parts \citep{AlMoulla+2022,Siegel+2022} formed at certain temperature ranges in the stellar photosphere. Contrary to the LBL method whose feasibility only relies on observational data, this \acrfull*{PBP} method also requires a priori knowledge about where certain spectral features are formed in the photosphere, which can be retrieved from synthetic spectra and thereafter mapped onto cross-matched lines in the observed spectra.

\subsection{Spectral synthesis}\label{Sect:3.1}

For the spectral synthesis, we made use of \texttt{PySME} \citep{Wehrhahn+2023}, a Python version of \acrlong*{SME} \citep[\texttt{SME};][]{Valenti&Piskunov1996,Piskunov&Valenti2017}. For each target we used a unique line list from the \texttt{VALD}3\footnote{\texttt{VALD} stands for \acrlong*{VALD}.}$^{,}$\footnote{\url{http://vald.astro.uu.se}} database \citep{Piskunov+1995,Kupka+2000,Ryabchikova+2015}, which were queried for a wavelength interval of $\SI{3500}{}$--$\SI{7000}{\angstrom}$, slightly larger than the range of the HARPS spectrograph. In the queries we used the stellar parameters in Table~\ref{Tab:01} and a normalized line depth threshold of $0.01$ for all targets except the M dwarf (Gl 229) for which we used a threshold of $0.15$ in order to produce line lists with less than $\num{100000}$ lines (the maximal output of \texttt{VALD}). We used a grid of \texttt{MARCS} \citep{Gustafsson+2008} model atmospheres interpolated for the same stellar parameters as for the queries. The syntheses were broadened with their adopted $v\sin{i}$ (see Table~\ref{Tab:02}), and micro- and macroturbulences adopted from \citet{Valenti&Fischer2005}. The microturbulence was fixed to
\begin{equation}\label{Eq:3}
v_{\mathrm{mic}} = \SI{0.85}{\kilo\meter\per\second}
\end{equation}
for all stars, and the macroturbulence was set to
\begin{equation}\label{Eq:4}
v_{\mathrm{mac}} = \left(3.98 + \frac{T_{\mathrm{eff}} - \SI{5770}{K}}{\SI{650}{K}}\right) \SI{}{\kilo\meter\per\second}\,.
\end{equation}
Note that in Eq.~\ref{Eq:4}, we have changed the sign of the $T_\mathrm{eff}$-dependent term compared to \citet{Valenti&Fischer2005}, since their equation has an apparent typo. This is evident from the plotted relation in their Fig.~3 and was also pointed out by \citet{Takeda&UeNo2017}. Although the relations in \citet{Valenti&Fischer2005} were derived for FGK stars, we assume they approximately extend to early M dwarfs as well. Furthermore, minor discrepancies in turbulence broadening is not a limiting factor for our analysis since the subsequent rejection of poorly synthesized lines is relative for each star, hence if all lines are equally over- or under-broadened the final selection remains almost identical. The spectra were finally convolved with a Gaussian instrumental profile to resemble the HARPS resolution of $R\,{=}\,\num{115000}$. For each target, a second synthesis excluding rotational, turbulent and instrumental broadening was computed for the identification of line blends not otherwise seen.

\subsection{Average formation temperature}\label{Sect:3.2}

The average formation temperature, denoted $T_{1/2}$, was then calculated at each sampled wavelength point of these synthetic spectra by integrating the flux contribution function and inferring the value in the temperature grid at which it reached 50\% of its maximum value (see \cite{AlMoulla+2022} for further details). Although this value follows the flux variations on a line-by-line basis, it does not hold globally, and thus lines of equal flux depth do not necessarily form at the same average temperature subject to their atomic properties as well as the possibly different level of continuous absorption locally around them.

\begin{figure*}[t!]
	\includegraphics[width=\textwidth]{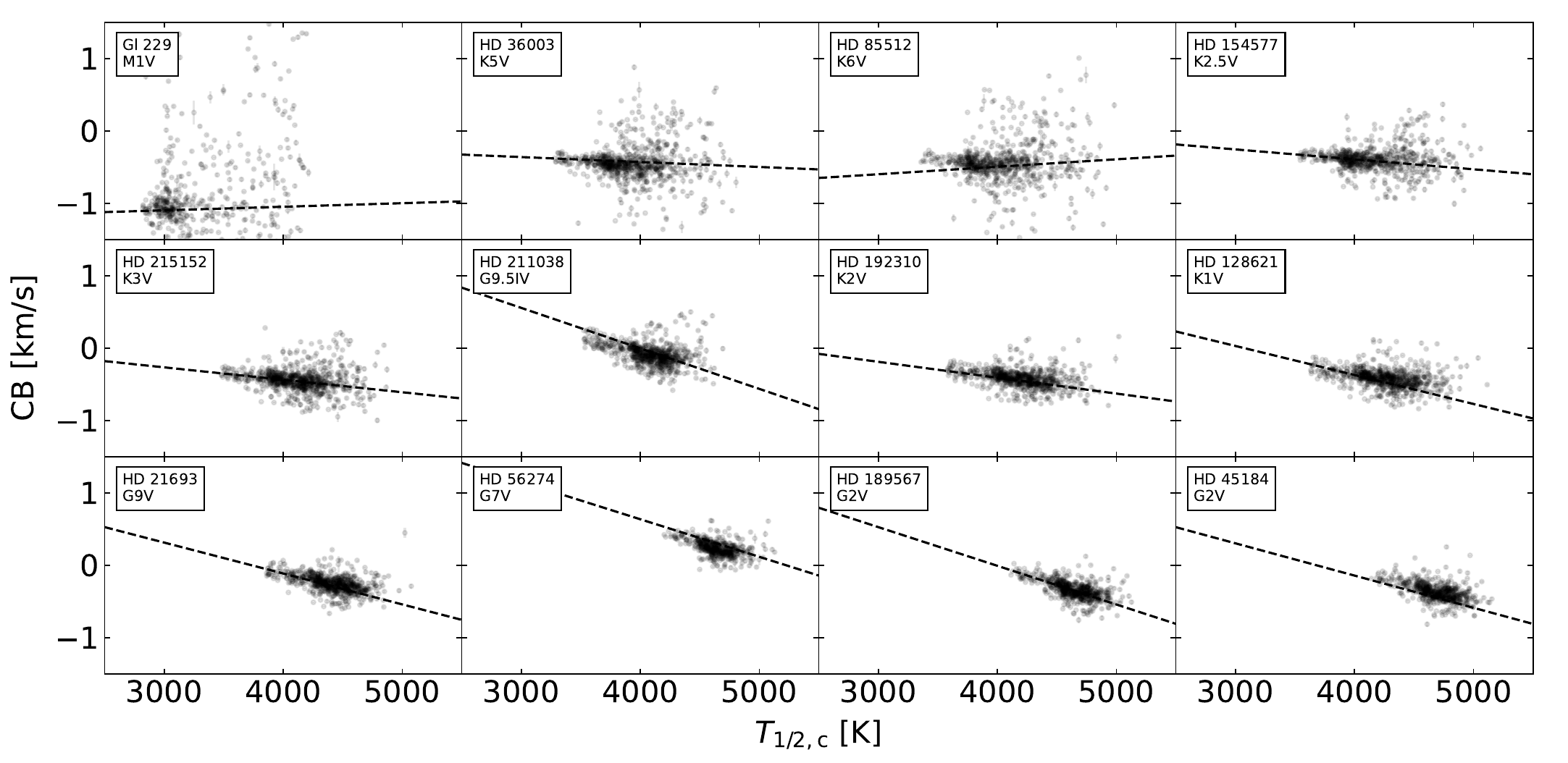}
	\caption{CB as a function of formation temperature at the line cores, $T_{1/2,\mathrm{c}}$, for the stellar sample. The stars and their spectral types are labeled in the upper left corner of each subplot, and ordered by effective temperature, from left to right and top to bottom. For most stars, the CB shows a linear trend with $T_{1/2,\mathrm{c}}$, indicating that lines formed at higher temperatures, i.e., deeper layers of the photosphere where convective motion is stronger, are subject to larger velocity shifts. The individual spectral lines are shown as points, and the fitted linear trends are shown as dashed lines.}
	\label{Fig:1}
\end{figure*}

\begin{figure}[t!]
	\includegraphics[width=\linewidth]{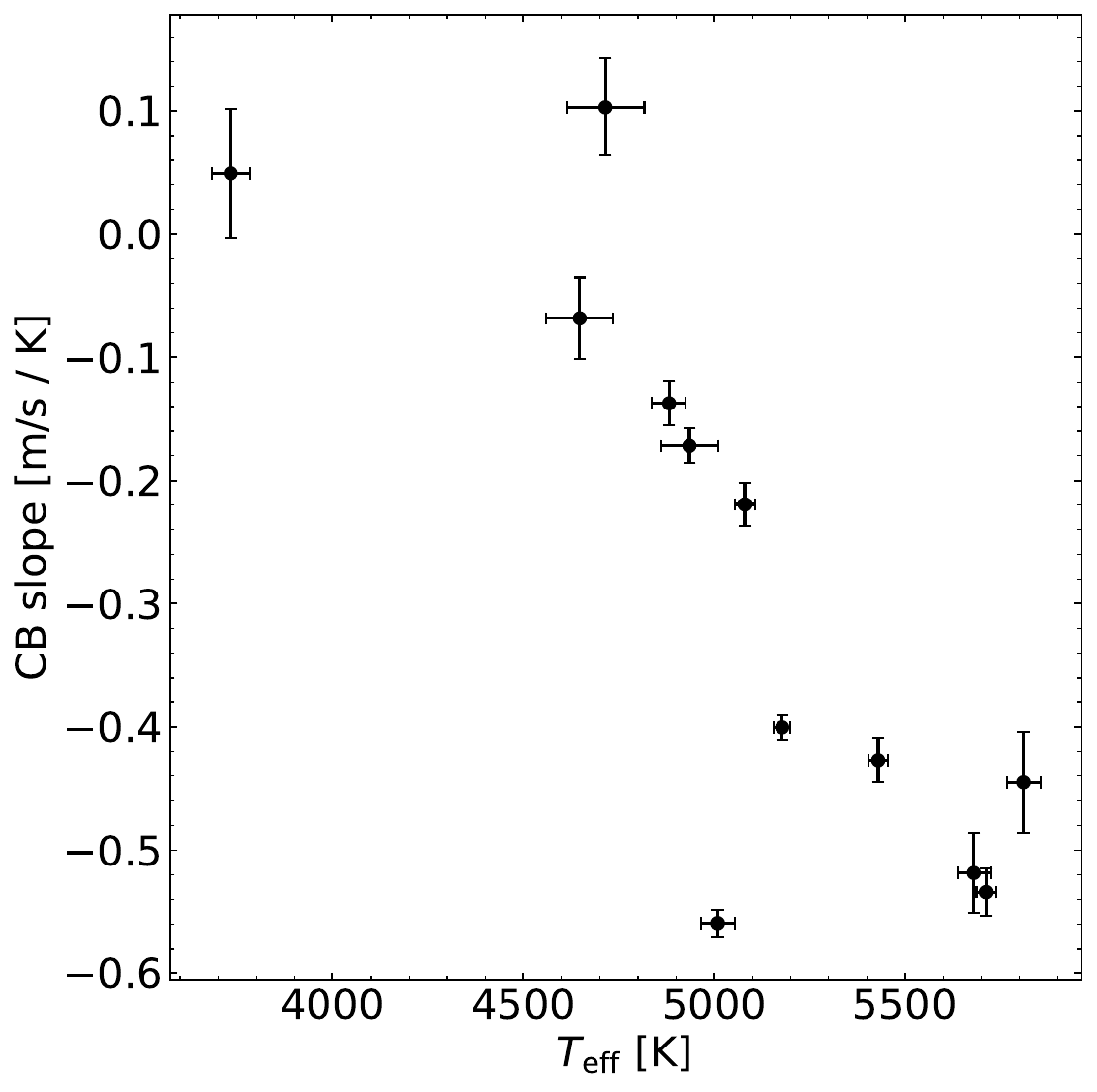}
	\caption{Slope of the CB with respect to the average formation temperature of the line cores, as measured from the linear fits seen in Fig.~\ref{Fig:1}. The slope shows a decrease with the stellar effective temperature.}
	\label{Fig:2}
\end{figure}

\subsection{Convective blueshift}\label{Sect:3.3}

For each observed line in the master spectra (see Sect.~\ref{Sect:2.1}), defined as a local minimum bounded by two local maxima, we re-calibrated the central wavelength by fitting a second-order polynomial to the central-most seven points. Once the formation temperature spectra were computed, we interpolated their value at the line center, denoted $T_{1/2,\mathrm{c}}$, of each observed line. We then identified the species of the observed lines using \texttt{VALD} line lists. The cross-match with \texttt{VALD} was done by querying a \texttt{VALD} line list according to the stellar parameters of each star, and then finding all \texttt{VALD} lines which fall within the left and right borders of the observed lines. For observed lines which only have one possible match in the \texttt{VALD} list, the \texttt{VALD} line was taken as the correct identification. Using these uniquely identified lines, an average wavelength dispersion (caused by uncorrected shifts and/or discrepancy in the wavelength solution) was measured. For observed lines which have multiple possible matches, the \texttt{VALD} line with the smallest Euclidean norm of the wavelength and depth differences was adopted if the wavelength was within the dispersion measured on uniquely identified lines. Some observed lines were expected to remain unidentified due to having poorly characterized atomic properties, being tellurics or noisy patterns falsely flagged as stellar lines. We thereafter computed the Doppler shift between the observed central wavelengths of identified lines and the restframe wavelengths from \texttt{VALD}, known as the \acrfull*{CB}. This shift originates due to the motion of convective surface layer in late-type stars, where ascending granules emit more light and cover a larger area than the descending intergranular lanes, causing a net blueshift \citep{Dravins+1981}. In order to calibrate the CB to an absolute value, we also subtracted a constant gravitational redshift for each star,
\begin{equation}\label{Eq:5}
    v = \frac{GM}{Rc}\,,
\end{equation}
where $G$ is the gravitational constant, and $c$ is the speed of light in vacuum. We note that the uncertainty on the gravitational redshift is not propagated into the individual line CBs, as the gravitational redshift is simply a constant offset for each star which we consider in order to have most of our values being negative and hence truly blueshifts. For the remainder of our analysis we only focus on the CB slopes and not the zero-points.

After rejecting blended lines (as determined both from the master spectra themselves and from the syntheses void of any broadening other than thermal) and poorly synthesized lines (see \cite{AlMoulla+2022} for the detailed applied criteria), we were left with our final selection of lines, presented in Fig.~\ref{Fig:1}. We remark that, for most stars, the CB decreases (to larger absolute values) with average formation temperature in the line core as a result of probing hotter photospheric regions where the convective motions are more turbulent and hence producing larger blueshifts. For each star, we regressed a linear model and present the best-fitting slopes in Fig.~\ref{Fig:2}, where the distribution of slopes appears to also follow a linear relation with effective temperature. The negative slope can be explained by the larger extent of the convective zone of later type stars, where turn-over timescales are longer and thus the turbulent motions expected to be slower. The direction of this trend is also in agreement with the findings of \cite{Liebing+2021}, who fitted a scaling factor to the solar CB vs. line depth relation for 810 F-M stars and found the scaling factor to decrease with effective temperature.

From our limited sample, it remains unclear whether the CB slope plateaus to zero for the coolest stars, or becomes positive. Given the distribution of points in Fig.~\ref{Fig:1}, the linear fits for Gl 229 (the M dwarf) and HD 85512 appear to be poorly constrained. For these stars, the spectral syntheses had larger discrepancies with observations compared to the earlier type stars, and thus less reliable derived formation temperatures. Another outlier to be remarked is HD 211038 which has a steeper slope compared to the stars in the vicinity of its effective temperature. HD 211038 is the only subgiant in our sample of otherwise \acrfull*{MS} stars, and it has a significantly lower surface gravity, $\log{g}{=}3.87{\pm}0.06$, compared to all the other stars. The steeper slope, indicating a stronger convection strength compared to MS stars of similar effective temperature, is in agreement with the empirical relations found by \cite{Liebing+2023} for evolved stars.

\subsection{Temperature-binned RVs}\label{Sect:3.4}

For this study we decided to divide the entire range of formation temperatures per star into intervals of equal sizes, denoted ${\Delta}T_{1/2}(N,n)$, where $N$ is the number of intervals and $n\,{=}\,[1,N]$ is the interval index going from cooler to hotter temperatures. We considered the $N\,{=}\,1$ (equivalent to no temperature binning, i.e., conventional LBL) and $N\,{=}\,4$ cases, in order to have comparable results with \cite{AlMoulla+2022}. It should be noted that the bounds of the temperature intervals for different stars do not necessarily agree with each other, since the intervals were assessed relative to the line-forming temperature regime of each star, i.e., the available range of temperatures as given by the spectral syntheses.

The PBP RV for a formation temperature interval was computed by template-matching segments of the individual observed lines which happen to fall within the considered temperature interval with the same entire line in the master spectrum \citep{Dumusque2018}. This was then repeated for each temperature interval, spectral line and spectrum in the time series. After an iterative $4\sigma$ outlier rejection on the RMS of each interval and line, we computed a weighted average across all remaining lines, resulting in a single RV time series per temperature interval. We require a minimum of 3 points for the template-matching, however, 3 points alone will produce a large uncertainty for a given line and temperature interval, such that it will likely be sigma-clipped when computing the line-averaged RV per temperature interval. Furthermore, the temperature intervals will not necessarily be similar across the wavelength range. The primary reason for this is the influence of continuous absorption on the derived formation temperature (see Fig.~2 in \citet{AlMoulla+2022}). In the optical range for Sun-like stars, the continuous absorption is dominated by the negative hydrogen ion, which increases its opacity up to about $\SI{1}{\micro\meter}$ \citep{Gray2008}. Therefore, for the HARPS wavelength range, bluer wavelengths are biased to hotter temperatures and redder wavelengths to cooler temperatures (because the increased opacity allows us to probe less of the photosphere). But it is exactly these kind if nuanced differences that we aim to achieve with our analysis.

The $N\,{=}\,1$ and $N\,{=}\,4$ RV time series are presented in Figs.~\ref{Fig:A1}--\ref{Fig:A12}, together with the nightly-averaged $S$ indices and their respective \acrlong*{GLS} \citep[GLS;][]{Lomb1976,Scargle1982,Zechmeister&Kurster2009} periodograms. The $S$ indices are retrieved from \texttt{DACE}\footnote{\texttt{DACE} stands for \acrlong*{DACE}.}$^{,}$\footnote{\url{https://dace.unige.ch}}, which in turn archives them from the HARPS \acrfull*{DRS}. The HARPS DRS computes the $S$ index in an equivalent way to the original Mount Wilson survey \citep[see][]{Lovis+2011}.

\begin{figure*}[t!]
	\includegraphics[width=\textwidth]{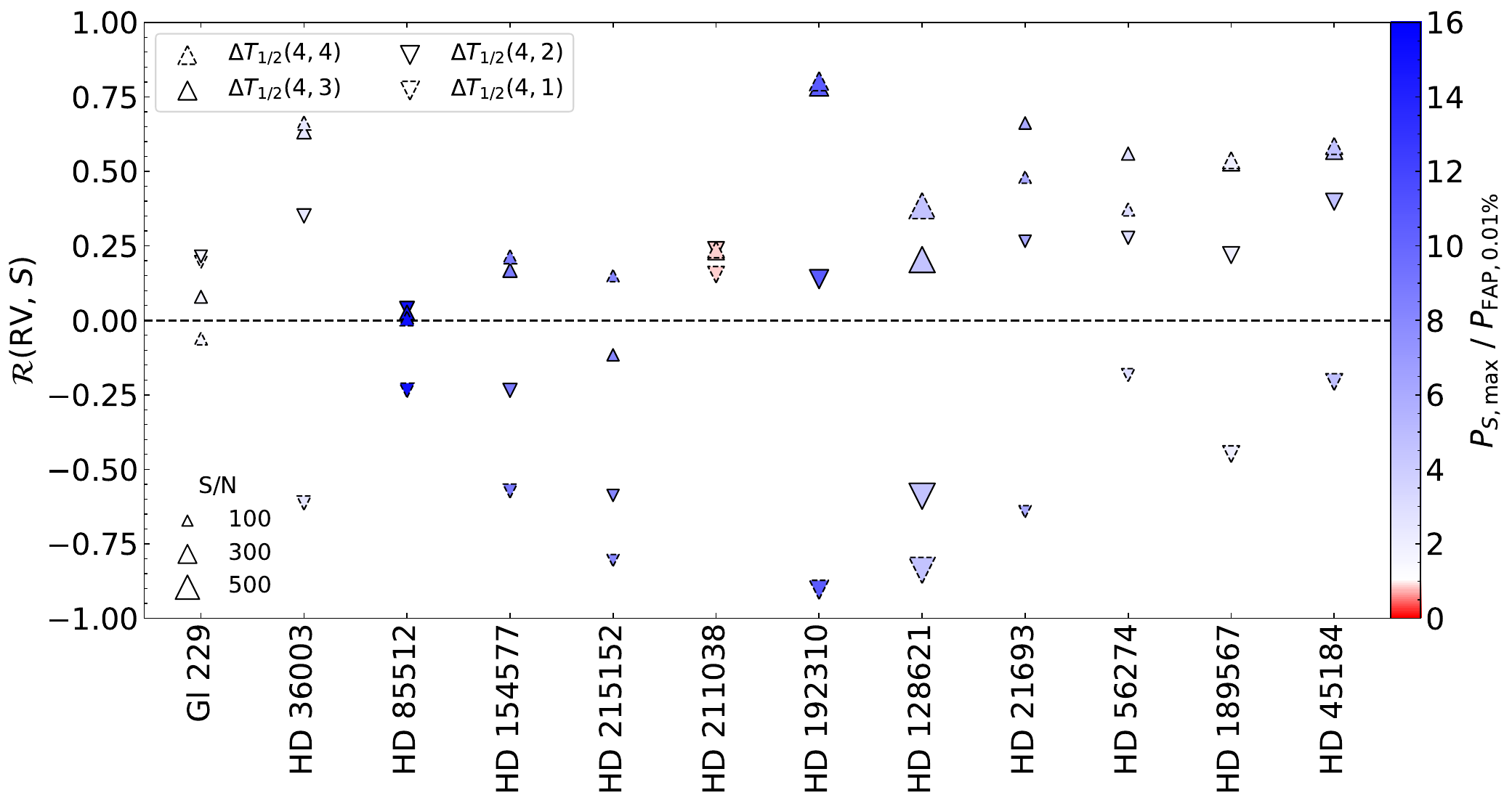}
	\caption{Pearson correlation, $\mathcal{R}$, between the temperature-binned RVs and the $S$ index. The stars are ordered by effective temperature. The marker size indicates the median S/N of the studied spectra of each star, and the marker color indicates the power, $P$, of the largest periodogram peak of the $S$ index relative to its $0.01$\% FAP level. The color map inverts at $1$, such that green points represent periodograms peaks above the $0.01$\% FAP, and red points represent peaks below it.}
	\label{Fig:3}
\end{figure*}

\section{Activity correlations of temperature-binned RVs}\label{Sect:4}

We compute the Pearson correlation coefficient between the $S$ index and each temperature-binned RV time series, which are shown in Fig.~\ref{Fig:3}. We find that for most stars, the RVs of line parts formed in hotter temperatures tend to be more positively correlated, and the RVs of line parts formed in cooler temperatures tend to be more negatively correlated. Furthermore, this relation appears to be amplified with the significance of the signal, as shown both by the median S/N of the stellar spectra for each star and by the power of the largest periodogram peak for the $S$ index relative to its $0.01$\% \acrfull*{FAP} level. The color-coded variable in Fig.~\ref{Fig:3} was selected as a quantification of how statistically significant the $S$ index variations are. We believe that this is better than stating a potentially inaccurate estimate of the $S$ index semi-amplitude since the cycles are not fully sampled. We find no clear dependence with effective temperature.

These correlations are in agreement with the findings of \citet{AlMoulla+2022} (see their Figs.~A.1 and B.3). However, while they attributed the measured signal at hotter temperature ranges to the suppression of CB by faculae, using simulations of the Sun with \texttt{SOAP-GPU} \citep{Zhao&Dumusque2023}, no clear explanation has been established for the seemingly anti-correlated activity pattern observed at cooler temperatures. The $S$ index, derived from the deep \ion{Ca}{II} H and K lines, is known to exhibit chromospheric activity, however, the reason why activity signals at hotter to cooler formation temperatures transition from positively correlated, to weakly correlated, to negatively correlated with the $S$ index requires a more fundamental understanding of the types of active regions and the velocity-altering processes to which chromospheric activity is connected. Forthcoming results by \citet{Siegel+inprep} show that RVs measured on the top versus bottom parts of the CCF (approximately translating to spectral parts formed at hotter versus cooler temperature ranges) correlate with the photometric and convective influences of active regions, respectively. These results strongly indicate that spectral parts formed at different temperatures are primarily influenced by separate physical processes. Their connection to the $S$ index remain the topic of future investigations. In this study, we showcase the consistency of the observed differential correlation for magnetically active stars.

\begin{figure*}[t!]
    \includegraphics[width=\textwidth]{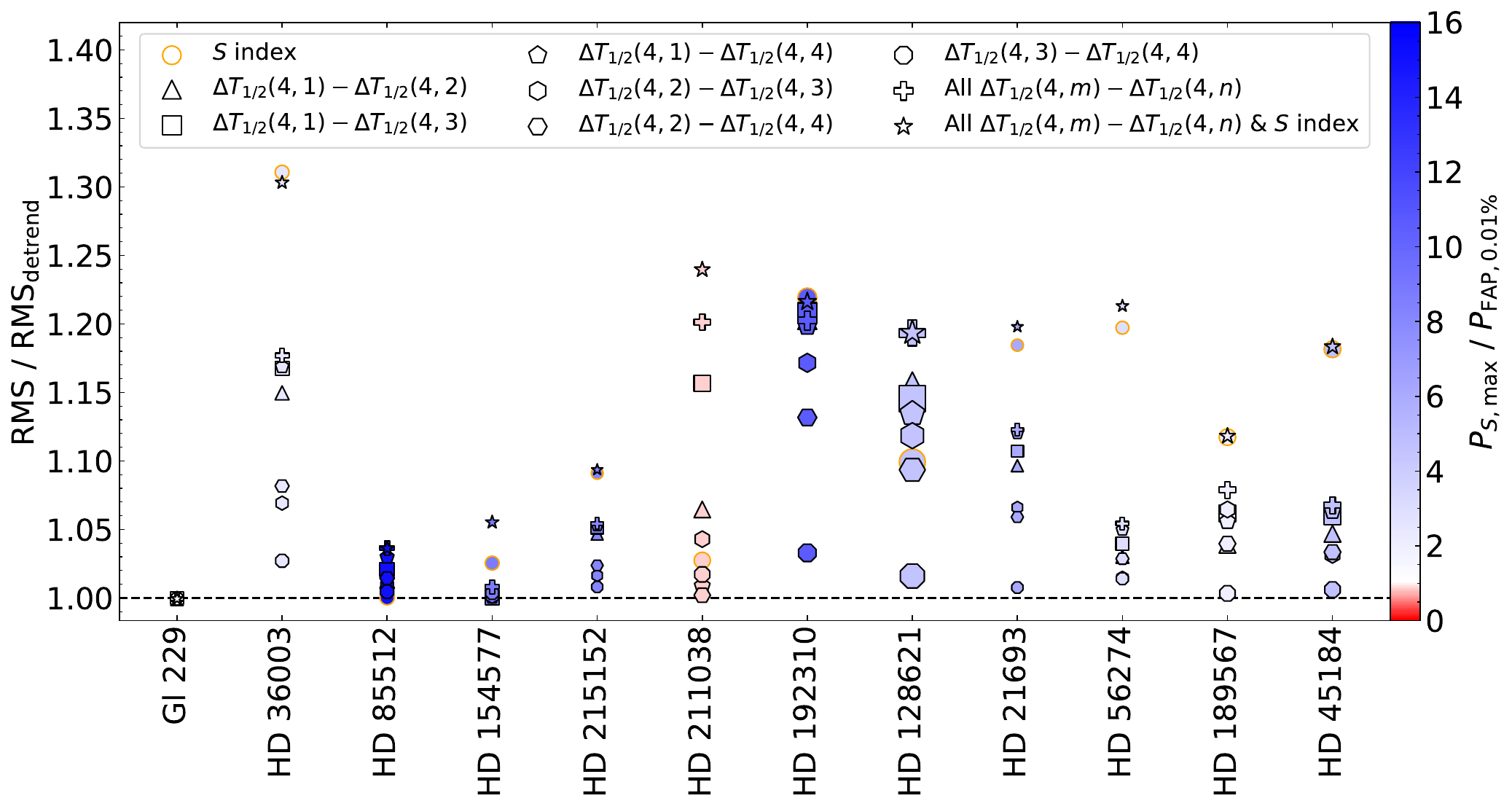}
	\caption{RMS improvement of the ${\Delta}T_{1/2}(1,1)$ RV time series after linearly detrending with different proxies. The markers represent detrending with either the $S$ index alone (circles with orange edges), different 4-bin residuals (convex polygons with black edges), a multi-linear combination of all 4-bin residuals (crosses), and a multi-linear combination of all 4-bin residuals and the $S$ index (stars). The marker sizes and colors indicate the same properties as described in Fig.~\ref{Fig:3}.}
	\label{Fig:4}
\end{figure*}

For a few stars in our sample (HD 215152, HD 192310 and HD 128621), we find a striking, near-perfect (${\leq}{-}0.8$) anti-correlation for the ${\Delta}T_{1/2}(4,1)$ time series. Due to the largely deviating covariances of these RVs, we decided to detrend the ${\Delta}T_{1/2}(1,1)$ time series (which contains all RV information) with proxies composed from differences of various combinations of the 4-bin RVs. The reason for taking the residual of two RV time series as an activity index and not the RV time series themselves is to cancel out the potential presence of a planetary signal in the proxy, since we expect pure Keplerian signals to be of equal amplitude regardless of the bin configuration. We compare the ratio of the RV \acrfull*{RMS} before and after detrending to assess the improvement factor of the RV precision. We try detrending with the $S$ index alone as a reference case, before we detrend with one RV residual at a time, followed by a \acrfull*{MLR} with all combinations of 4-bin residuals, and then a MLR with all 4-bin residuals and the $S$ index combined. The RMS improvement factors for each star and detrending method are shown in Fig.~\ref{Fig:4}. We find that for a majority of stars in our sample the MLR regression with all bin residuals and the $S$ index is able to decrease the RMS to a greater extent than the $S$ index alone, indicating that the 4-bin RV time series are likely tracing different activity contributions than those of the $S$ index.

\section{Discussion and conclusions}\label{Sect:5}

In \citet{AlMoulla+2022}, the authors demonstrated that for two MS stars for which we can reach extremely high S/N (${>}1000$ for the G2V Sun and the K1V $\alpha\,\mbox{Cen}\,\mbox{B}$), it was possible to precisely measure the RV of different layers in the stellar atmosphere and to confidently measure the differential effect of stellar variability as a function of line formation temperature, which is expected from our understanding of the RV activity signal \citep[e.g.,][]{Meunier+2017}. In this paper, we further extend the analysis to 12 stars, with spectral types from G2V to M1V and observations that have a more reasonable S/N (${\sim}100{-}300$). We show again the differential effect of stellar variability as a function of depth inside the stellar photosphere, which demonstrates that this technique can be used to probe and correct for stellar activity on a wide variety of MS that are observed at high, but not extreme, S/N.

To confidently measure the differential effect of stellar variability as a function of line formation temperature, there are many aspects to consider and justify. Carrying out the analysis for solar observations has the advantage of theoretical models, such as the spectral synthesis used to derive the formation temperature itself, usually being optimized for the Sun as a reference target. In addition, solar observations enjoy the benefit of rarely being noise limited due to their wealth of photons. Hence to extend the analysis to solar-like stars requires the assumption of precise enough spectral modeling for other stellar parameters, as well as efficient high-resolution spectrographs that can reach high signal-to-noise ratio for stellar observations.

In our work, we justify the use and accuracy of synthetic spectra mapped to high-resolution observations by verifying the consistency of derived relations, such as the net blueshift of line cores (see Sect.~\ref{Sect:3.3}). Selecting unblended lines with relatively small observation-synthesis flux residuals, we find a linear relation between the CB and core formation temperature analogous to the third signature of granulation observed when substituting temperature for line depth \citep{Gray2009,Reiners+2016,Liebing+2023}. This also allows us to determine how our synthesis becomes successively less reliable for cooler stars as the relation becomes less constrained (cf. Fig.~\ref{Fig:1}).

By nightly-stacking and correcting our spectra for instrumental systematics using \texttt{YARARA}, we are able to extract temperature-binned RVs to an extent where the intrinsic variations still significantly exceed the white noise. Some of the selected stars do have confirmed planetary companions, however, planets with RV semi-amplitudes larger than a few meters-per-second are removed by \texttt{YARARA}. Any potentially remaining lower-mass planets would not affect our results significantly since they would contribute negligibly to the RV RMS and not be detectable with our limited methodology even after the improvement of the MLR.

We emphasize again the opposing effect of stellar activity on line parts formed at hotter versus cooler photospheric temperatures, as seen in the correlation coefficients with the $S$ index. The RVs originating from the coolest 25\% appear to be the most sensitive to $S$ index variations, and the negativity of the correlation alludes that this temperature regime is more sensitive to chromospheric activity differing from the dominating inhibition of CB at hotter temperatures. After detrending the RV time series computed on the entire range of formation temperatures with proxies composed of residuals of the time series computed on 4 temperature bins, we found that a MLR of all bin residuals together with the $S$ index yielded an RMS improvement greater than detrending with the $S$ index alone for a majority of the studied stars. On average, the improvement in RV RMS is of the order of 20\% when using the MLR and $S$ index. However, when analysing the results of Fig.~\ref{Fig:4}, it is still challenging to understand why in some cases the RV RMS is strongly reduced, while in some other cases, the improvement is minimal. For Gl 229, the RMS values after detrending are noticeably unchanged; the ratios are not exactly $1$, but close enough to not be discernible. This is not too surprising, however, given the very low activity correlations of all its RV time series (see Fig.~\ref{Fig:A1}).

From the results of this study, the residuals of the temperature-binned RVs are demonstrated as excellent proxies to probe stellar activity effect, mainly magnetic cycles, and to differentiate planetary long-period signals from stellar ones. However, since the efficiency of those proxies to correct for stellar signal change from star to star, without a clear correlation with spectral type, it remains to be explored how the formalism of using line formation temperatures to extract PBP RVs can be further leveraged to attain RVs targeting certain flavors of stellar activity or which are optimally mitigated for planet searches.

\text{}\\
\footnotesize
\noindent
\textit{Acknowledgments.} We thank the anonymous referee for their valuable comments which improved the quality of our manuscript. This work has been carried out within the framework of the National Centre of Competence in Research PlanetS supported by the Swiss National Science Foundation under grants 51NF40\_182901 and 51NF40\_205606. The authors acknowledge the financial support of the SNSF. M.C. acknowledges the SNSF support under the Post-Doc Mobility grant P500PT\_211024. This project has received funding from the European Research Council (ERC) under the European Union’s Horizon 2020 research and innovation program (grant agreement SCORE No. 851555). This research has made use of the SIMBAD database, operated at CDS, Strasbourg, France. This work has made use of the VALD database, operated at Uppsala University, the Institute of Astronomy RAS in Moscow, and the University of Vienna.

\balance

\bibliographystyle{aa}
\bibliography{References}

\onecolumn
\begin{appendix}

\section{Temperature-dependent RV time series}\label{Sect:A}

\begin{figure*}[h!]
    \includegraphics[width=\textwidth]{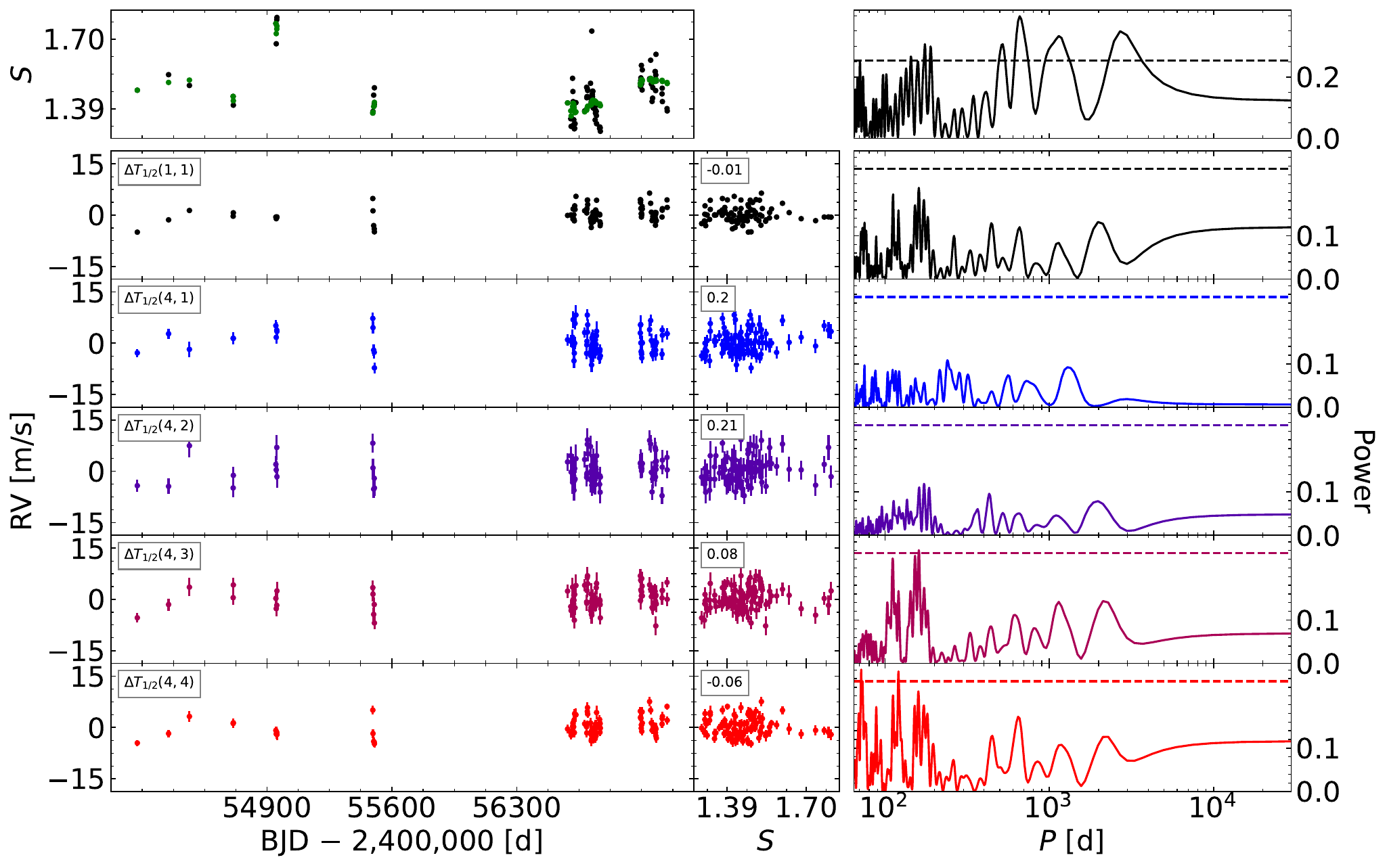}
	\caption{Time series and GLS periodograms of Gl 229. \textit{Left}: Time series of the $S$ index (first row), the RV computed for 1 formation temperature bin (second row), and the RV computed for 4 formation temperature bins (third to sixth row, in order of the coolest to hottest bin). For the $S$ index, black points show nightly-binned values and green points show nightly-binned values smoothed with a 200-day rolling average. \textit{Middle}: Correlations between the unsmoothed $S$ index and corresponding RVs from the left panels. The Pearson correlation coefficients, which are summarized in Fig.~\ref{Fig:3}, are shown in the upper left corners. \textit{Right}: GLS periodograms of the $S$ index and RVs from the left panels. The dashed lines show the corresponding $0.01$\% FAP levels.}
	\label{Fig:A1}
\end{figure*}

\begin{figure*}[h!]
	\includegraphics[width=\textwidth]{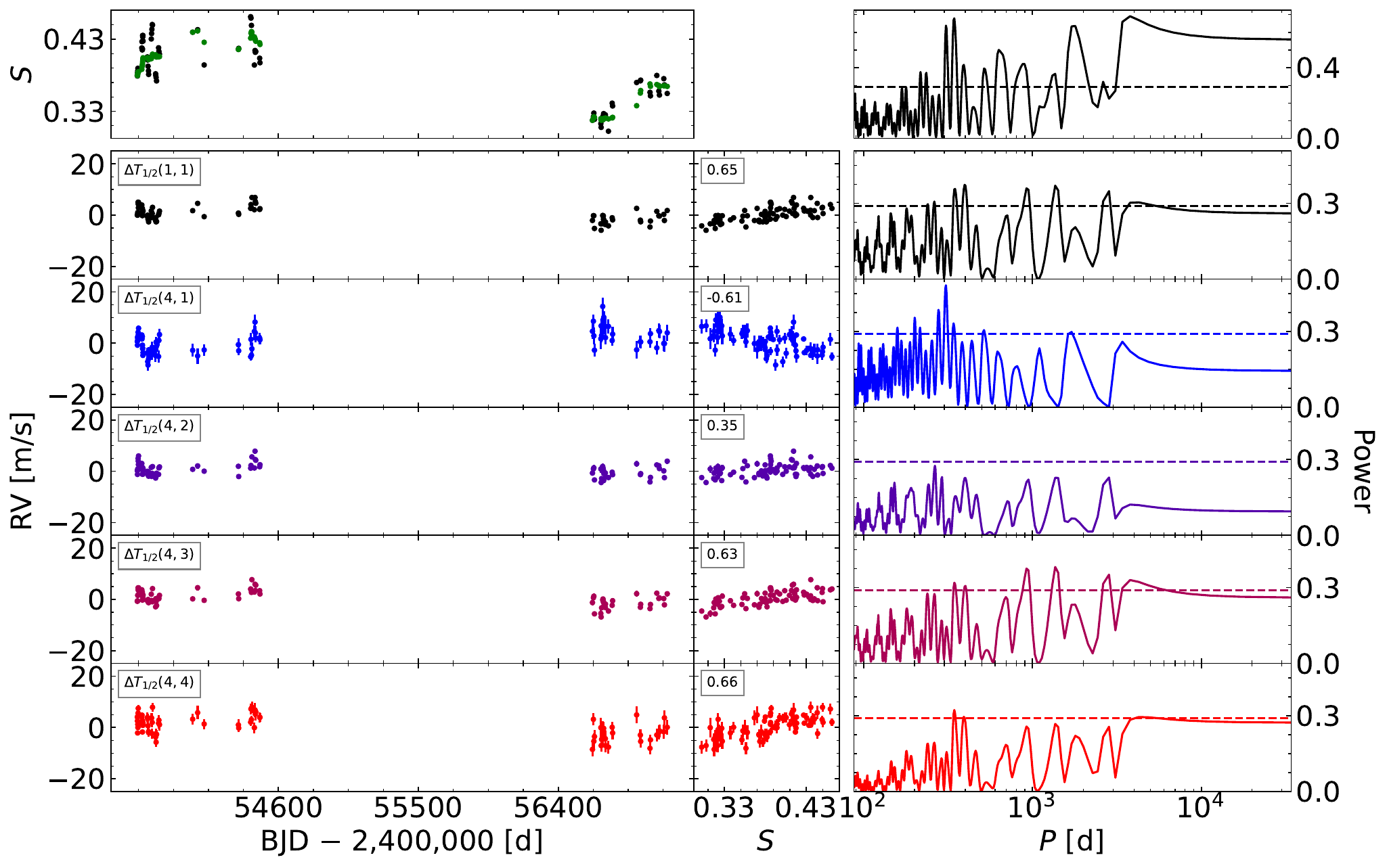}
	\caption{Same as Fig.~\ref{Fig:A1}, but for HD 36003.}
	\label{Fig:A2}
\end{figure*}

\begin{figure*}[h!]
	\includegraphics[width=\textwidth]{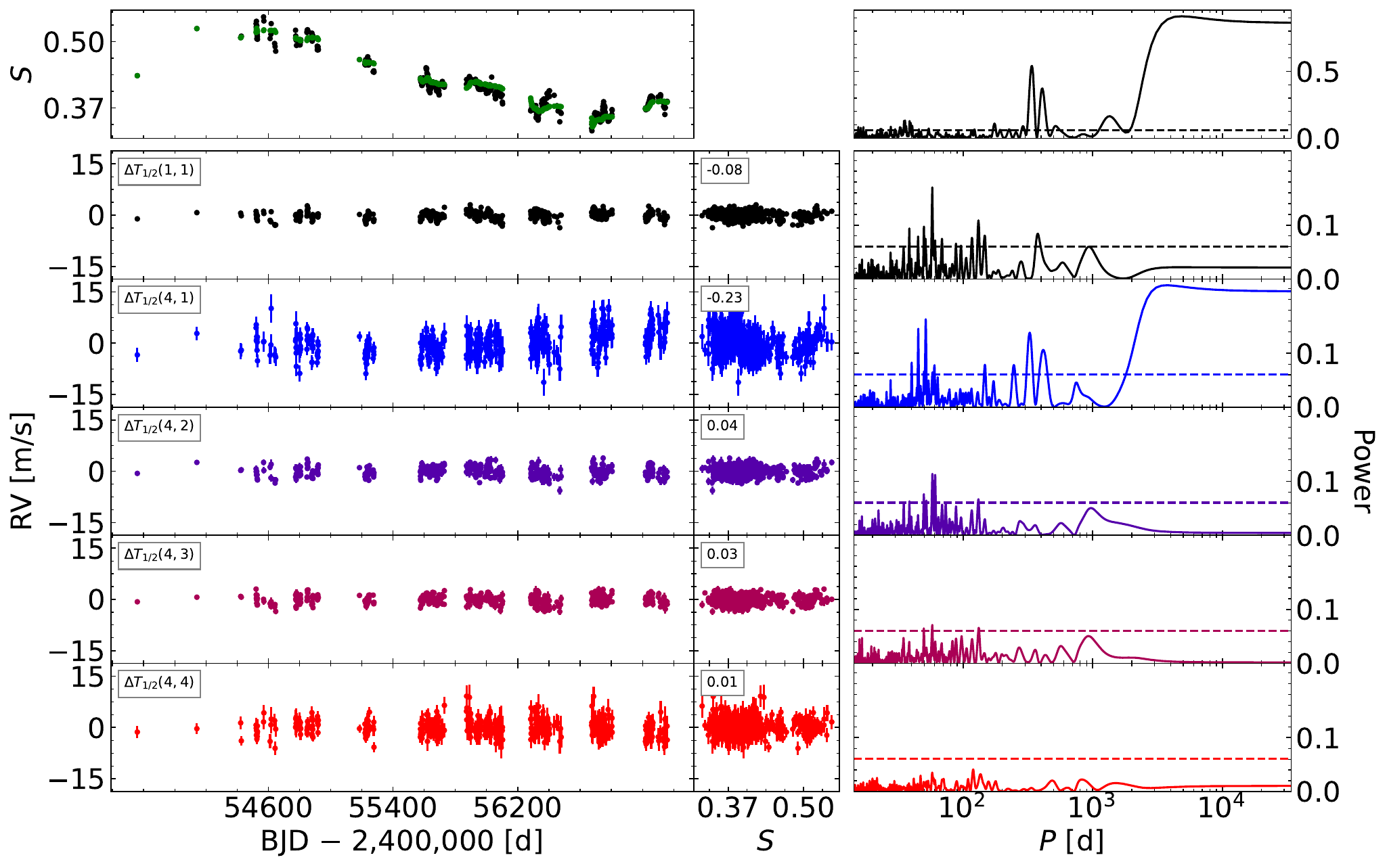}
	\caption{Same as Fig.~\ref{Fig:A1}, but for HD 85512.}
	\label{Fig:A3}
\end{figure*}

\begin{figure*}[h!]
	\includegraphics[width=\textwidth]{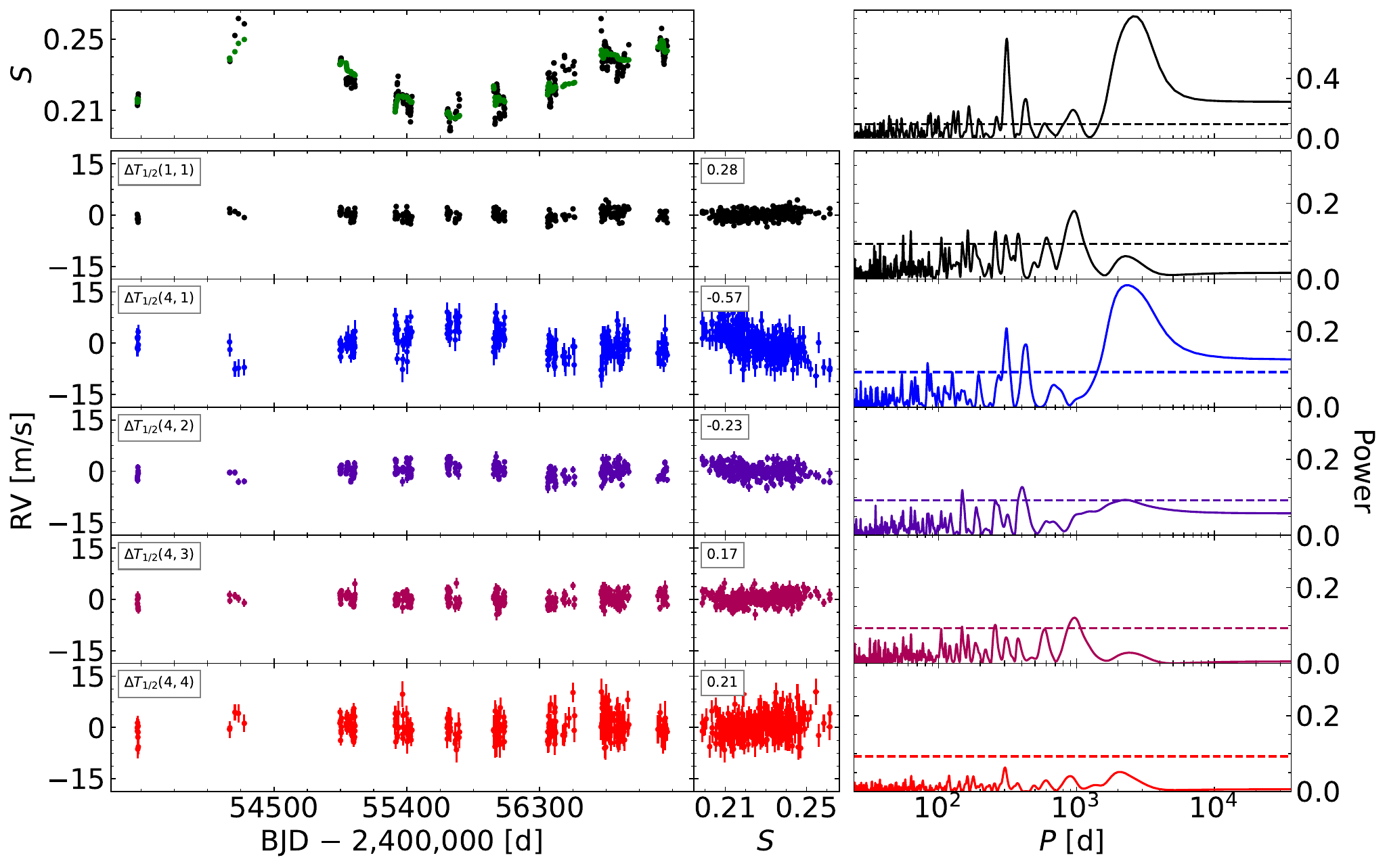}
	\caption{Same as Fig.~\ref{Fig:A1}, but for HD 154577.}
	\label{Fig:A4}
\end{figure*}

\begin{figure*}[h!]
	\includegraphics[width=\textwidth]{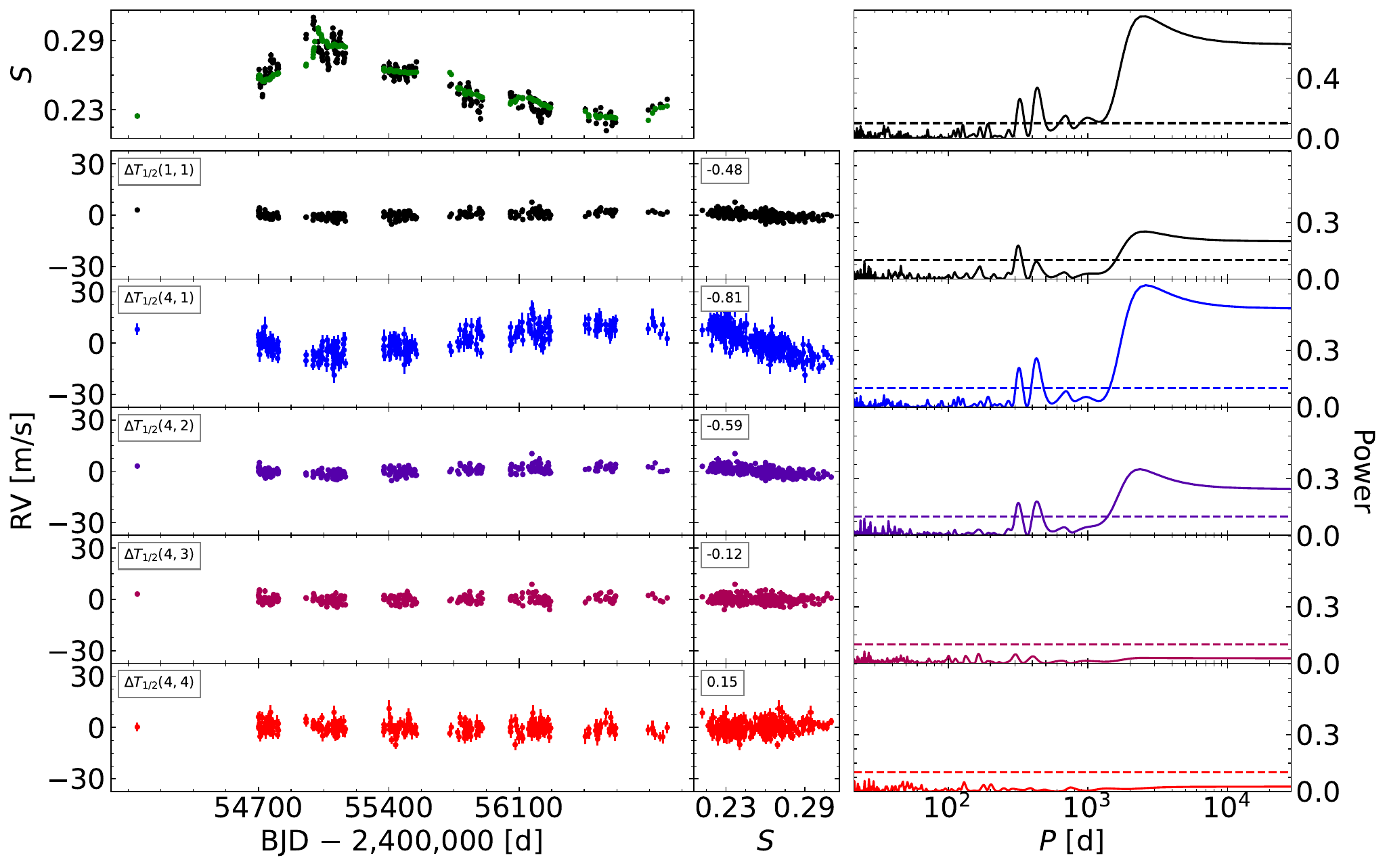}
	\caption{Same as Fig.~\ref{Fig:A1}, but for HD 215152.}
	\label{Fig:A5}
\end{figure*}

\begin{figure*}[h!]
	\includegraphics[width=\textwidth]{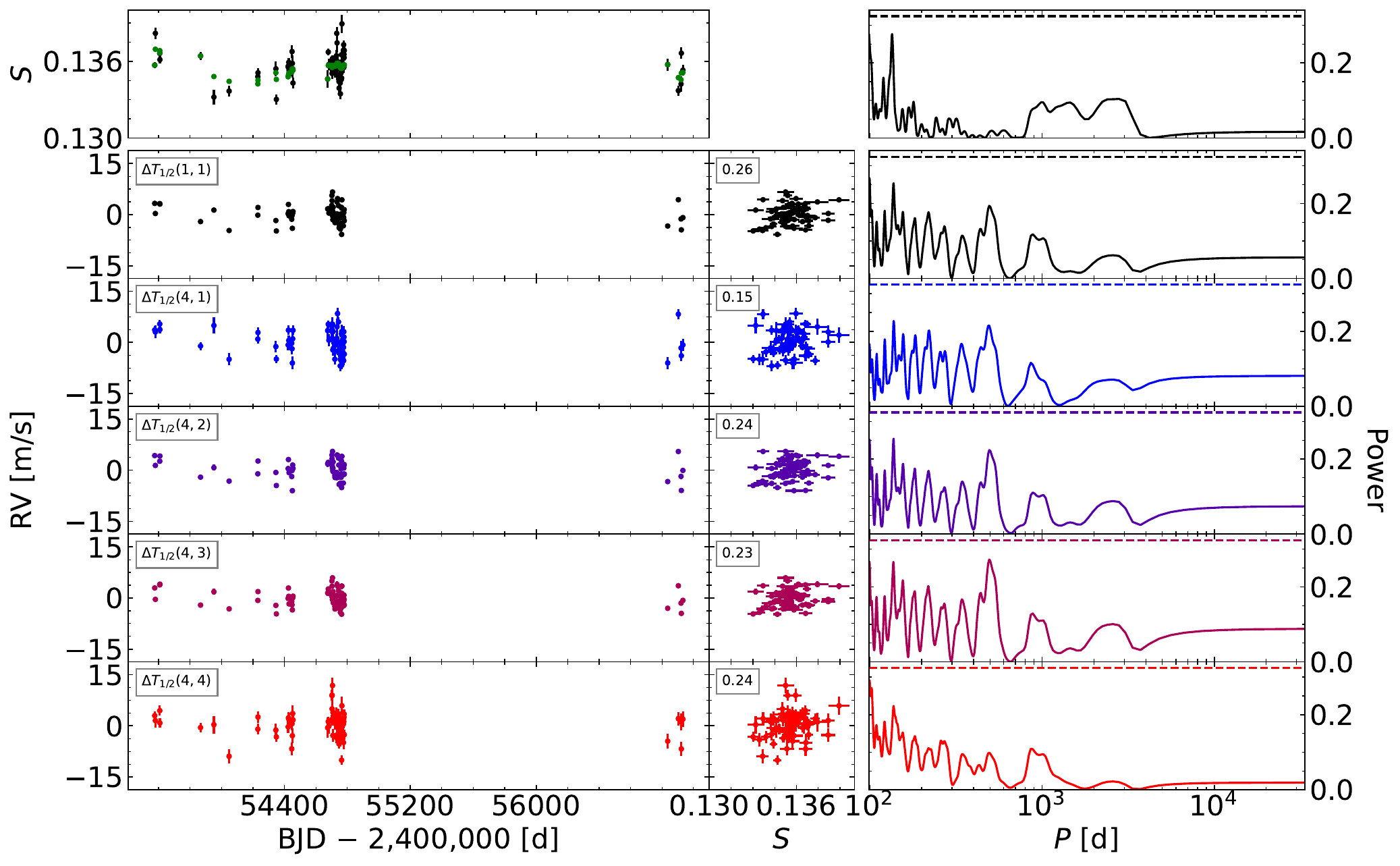}
	\caption{Same as Fig.~\ref{Fig:A1}, but for HD 211038.}
	\label{Fig:A6}
\end{figure*}

\begin{figure*}[h!]
	\includegraphics[width=\textwidth]{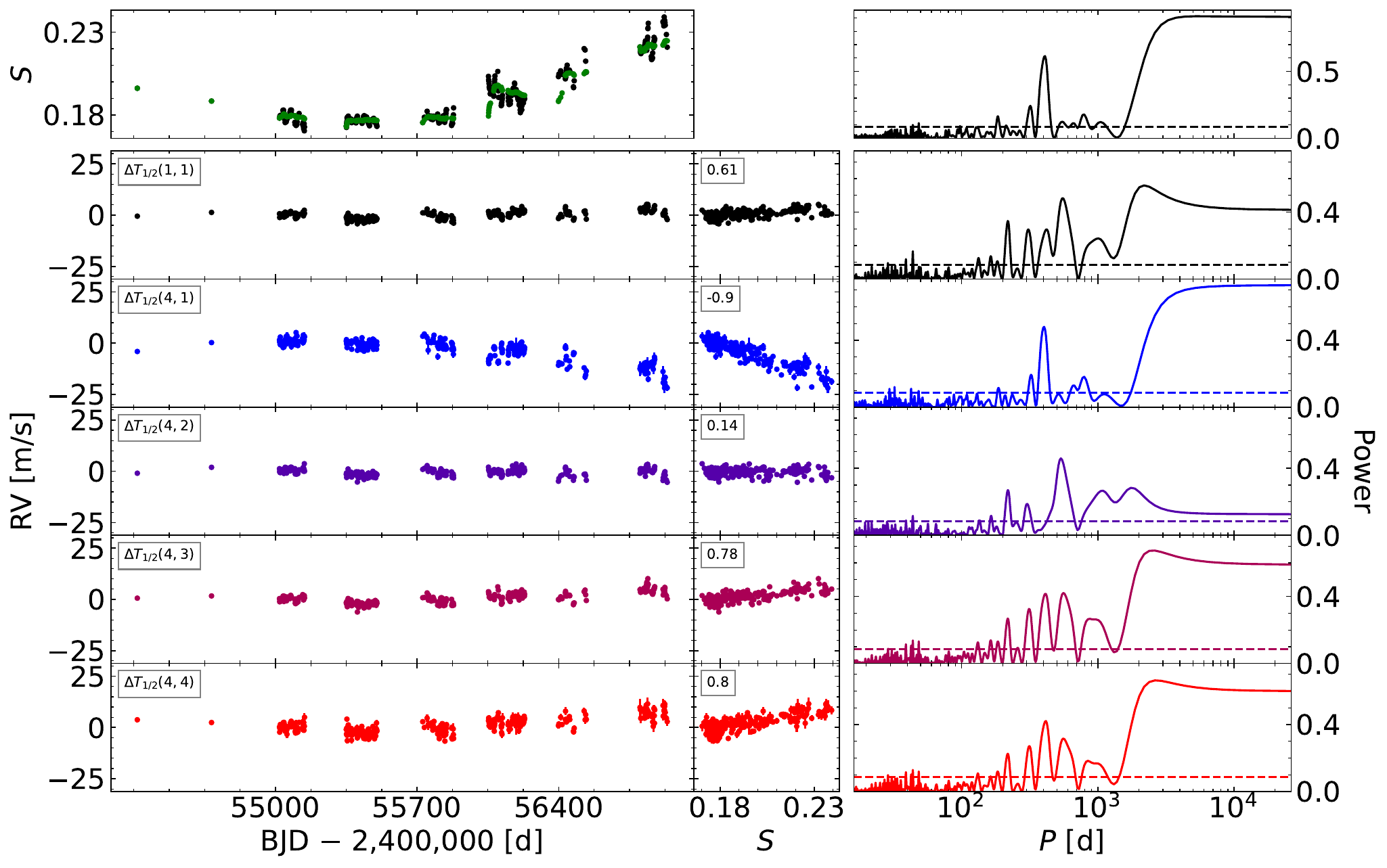}
	\caption{Same as Fig.~\ref{Fig:A1}, but for HD 192310.}
	\label{Fig:A7}
\end{figure*}

\begin{figure*}[h!]
	\includegraphics[width=\textwidth]{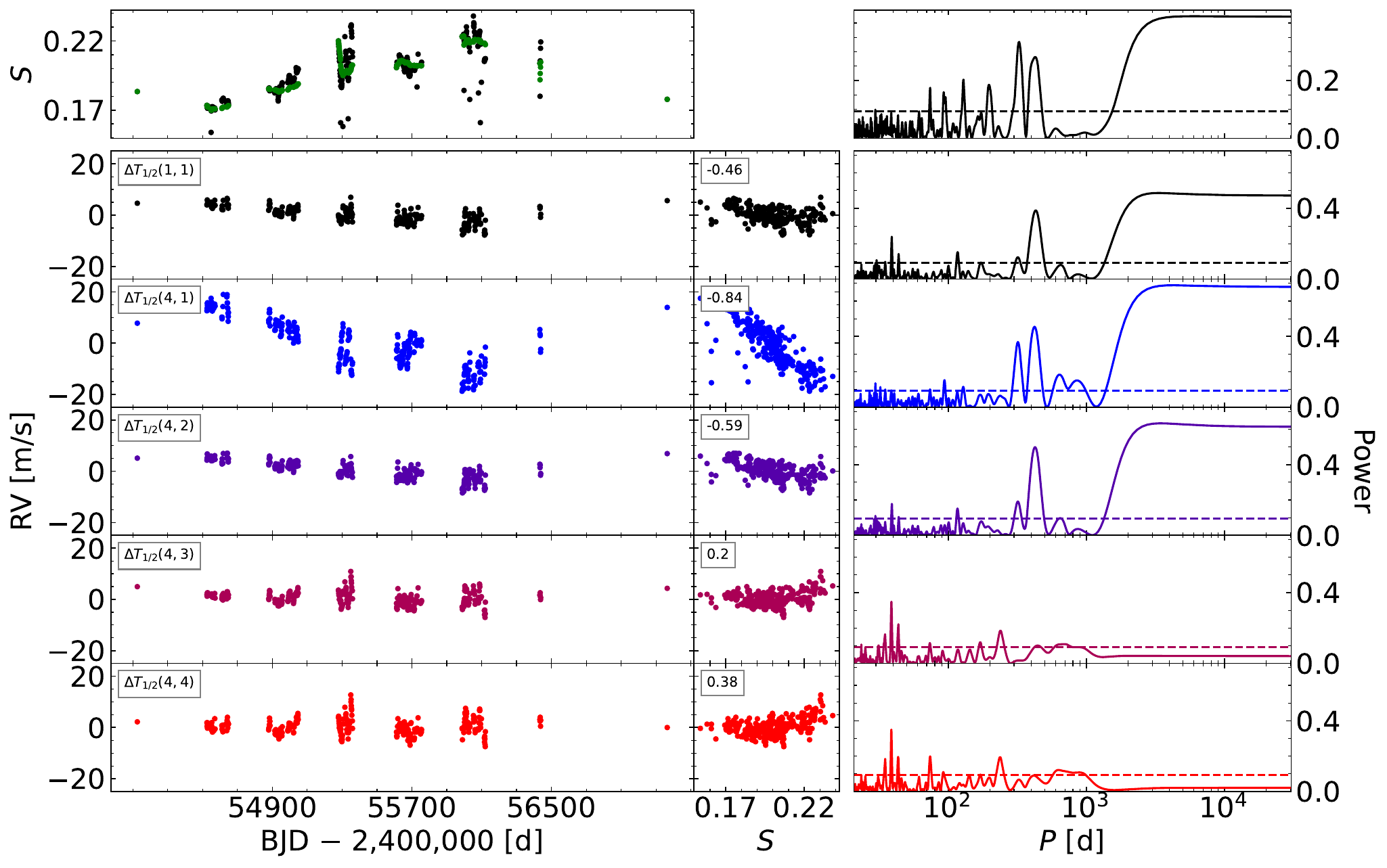}
	\caption{Same as Fig.~\ref{Fig:A1}, but for HD 128621.}
	\label{Fig:A8}
\end{figure*}

\begin{figure*}[h!]
	\includegraphics[width=\textwidth]{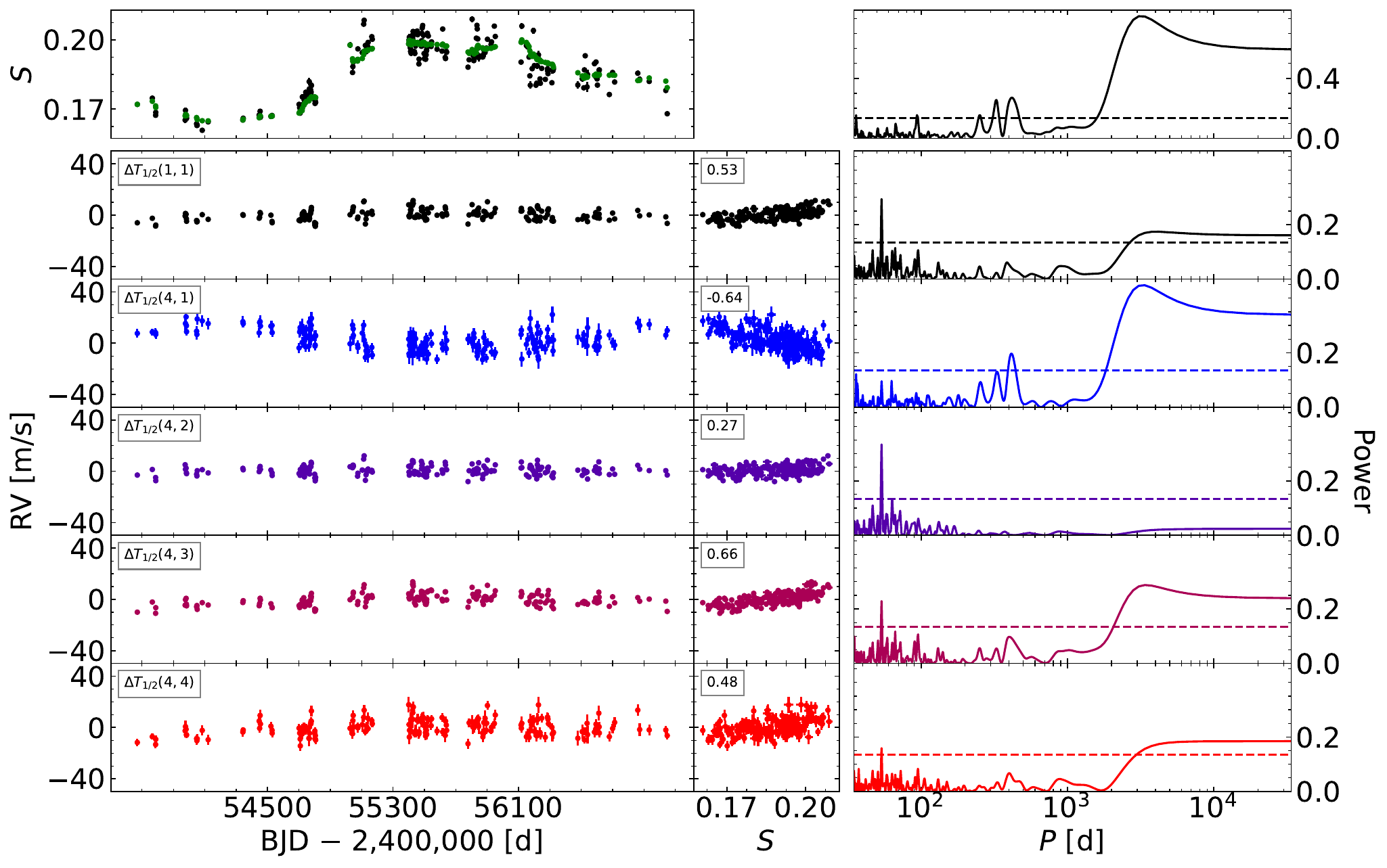}
	\caption{Same as Fig.~\ref{Fig:A1}, but for HD 21693.}
	\label{Fig:A9}
\end{figure*}

\begin{figure*}[h!]
	\includegraphics[width=\textwidth]{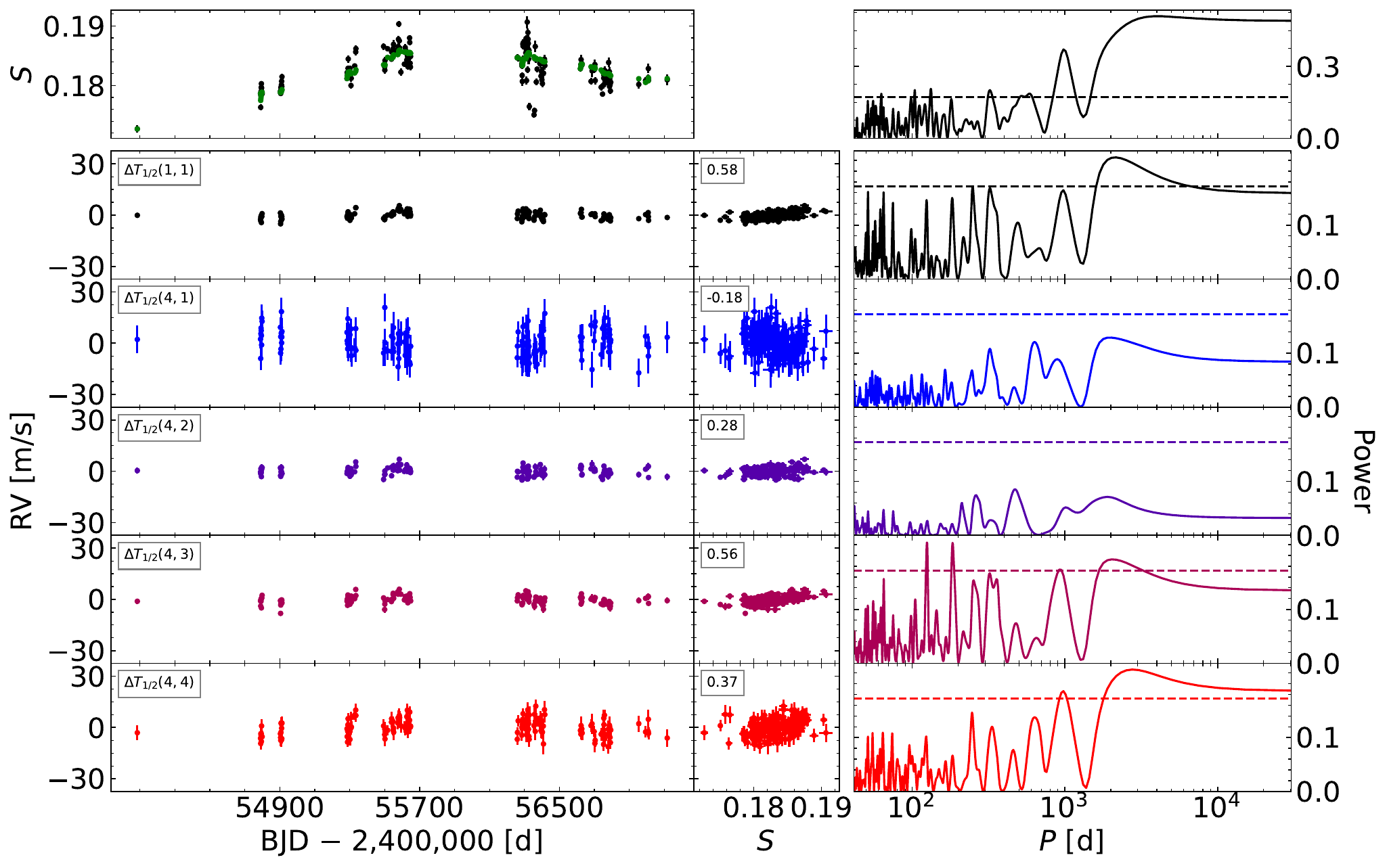}
	\caption{Same as Fig.~\ref{Fig:A1}, but for HD 56274.}
	\label{Fig:A10}
\end{figure*}

\begin{figure*}[h!]
	\includegraphics[width=\textwidth]{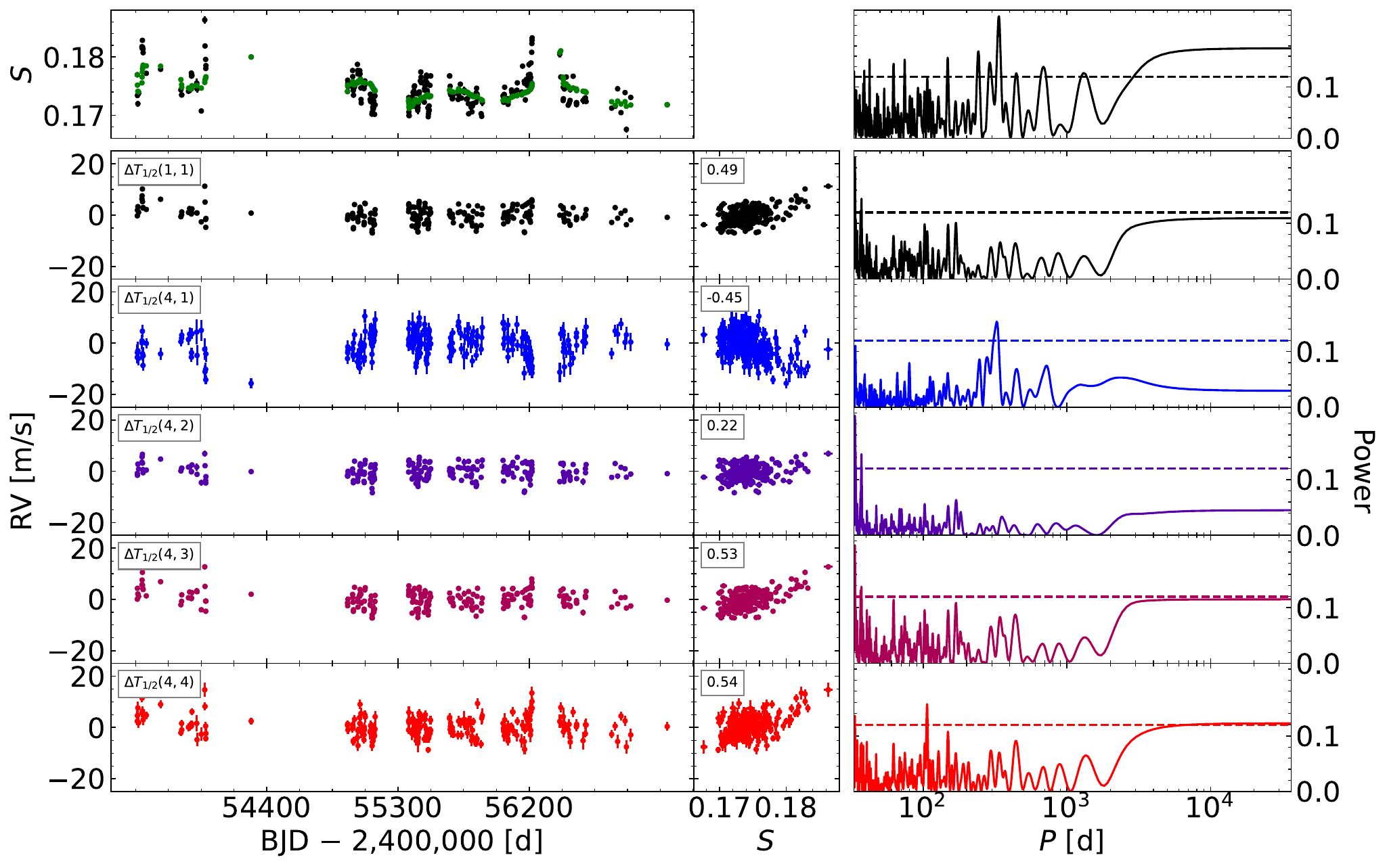}
	\caption{Same as Fig.~\ref{Fig:A1}, but for HD 189567.}
	\label{Fig:A11}
\end{figure*}

\begin{figure*}[h!]
	\includegraphics[width=\textwidth]{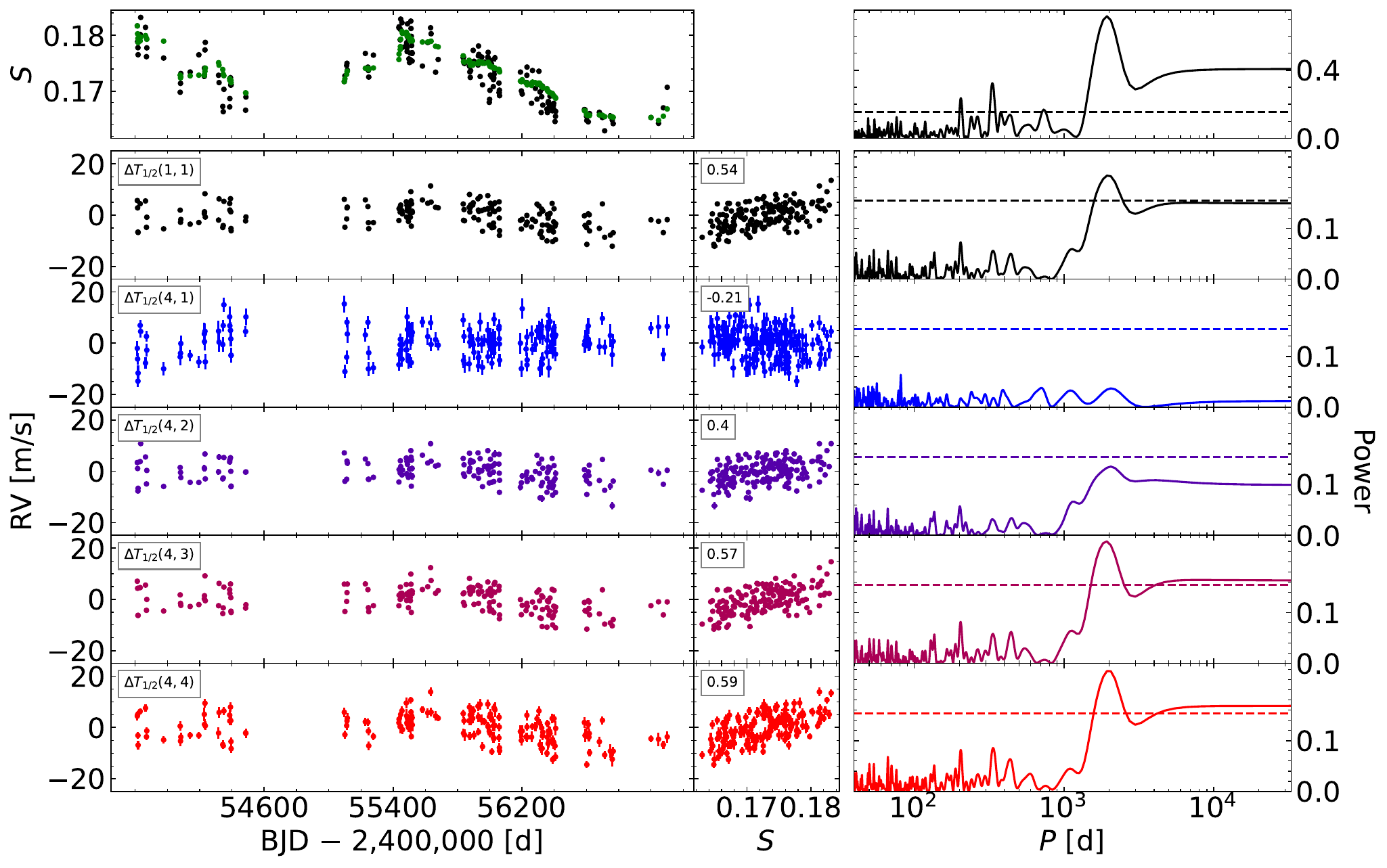}
	\caption{Same as Fig.~\ref{Fig:A1}, but for HD 45184.}
	\label{Fig:A12}
\end{figure*}

\end{appendix}

\end{document}